\documentclass[aps,prl,twocolumn,superscriptaddress,]{revtex4}
\usepackage{graphicx} 
\usepackage{bm}
\usepackage{color}
\usepackage{amssymb,amsmath} % erweiterte Captions
\usepackage{epstopdf}
\usepackage[utf8]{inputenc}
\usepackage{siunitx}
\usepackage{upgreek}
\usepackage{booktabs}
\usepackage{makecell}
\usepackage{natbib}
\usepackage{tabularx}

\usepackage[autolanguage, np]{numprint}

\begin{document} 
\pagenumbering{arabic}

\title{Magnetocrystalline anisotropy and exchange probed by high-field anomalous Hall effect in fully-compensated half-metallic Mn$_2$Ru$_x$Ga thin films}

\author{Ciar\'{a}n Fowley} \email[Corresponding author:] {c.fowley@hzdr.de}
\affiliation{Institute of Ion Beam Physics and Materials Research, Helmholtz-Zentrum Dresden - Rossendorf, Bautzner Landstra\ss e 400, 01328 Dresden, Germany}

\author{Karsten Rode}
\affiliation{AMBER and School of Physics, Trinity College Dublin, Dublin 2, Ireland}

\author{Yong-Chang Lau}
\affiliation{AMBER and School of Physics, Trinity College Dublin, Dublin 2, Ireland}

\author{Naganivetha Thiyagarajah}
\affiliation{AMBER and School of Physics, Trinity College Dublin, Dublin 2, Ireland}

\author{Davide Betto}
\affiliation{AMBER and School of Physics, Trinity College Dublin, Dublin 2, Ireland}

\author{Kiril Borisov}
\affiliation{AMBER and School of Physics, Trinity College Dublin, Dublin 2, Ireland}

\author{Gwena\"{e}l Atcheson}
\affiliation{AMBER and School of Physics, Trinity College Dublin, Dublin 2, Ireland}

\author{Erik Kampert}
\affiliation{Hochfeld-Magnetlabor Dresden (HLD-EMFL), Helmholtz-Zentrum Dresden - Rossendorf, Bautzner Landstra\ss e 400, 01328 Dresden, Germany}

\author{Zhaosheng Wang}
\altaffiliation{Current Address: High Magnetic Field Laboratory, Chinese Academy of Sciences, Hefei, Anhui 230031, People’s Republic of China}
\affiliation{Hochfeld-Magnetlabor Dresden (HLD-EMFL), Helmholtz-Zentrum Dresden - Rossendorf, Bautzner Landstra\ss e 400, 01328 Dresden, Germany}

\author{Ye Yuan}
\affiliation{Institute of Ion Beam Physics and Materials Research, Helmholtz-Zentrum Dresden - Rossendorf, Bautzner Landstra\ss e 400, 01328 Dresden, Germany}

\author{Shengqiang Zhou}
\affiliation{Institute of Ion Beam Physics and Materials Research, Helmholtz-Zentrum Dresden - Rossendorf, Bautzner Landstra\ss e 400, 01328 Dresden, Germany}

\author{J\"{u}rgen Lindner}
\affiliation{Institute of Ion Beam Physics and Materials Research, Helmholtz-Zentrum Dresden - Rossendorf, Bautzner Landstra\ss e 400, 01328 Dresden, Germany}

\author{Plamen Stamenov}
\affiliation{AMBER and School of Physics, Trinity College Dublin, Dublin 2, Ireland}

\author{J.M.D. Coey}
\affiliation{AMBER and School of Physics, Trinity College Dublin, Dublin 2, Ireland}

\author{Alina Maria Deac}
\affiliation{Institute of Ion Beam Physics and Materials Research, Helmholtz-Zentrum Dresden - Rossendorf, Bautzner Landstra\ss e 400, 01328 Dresden, Germany}
\date{\today}

\begin{abstract}
Magnetotransport is investigated in thin films of the half-metallic ferrimagnet Mn$_2$Ru$_x$Ga in pulsed magnetic fields of up to \SI{58}{\tesla}. 
A non-vanishing Hall signal is observed over a broad temperature range, spanning the compensation temperature ($\SI{155}{\kelvin})$, where the net magnetic moment is strictly zero, the anomalous Hall conductivity is \SI{6673}{\per\ohm\per\meter} and the coercivity exceeds \SI{9}{T}.
Molecular field modelling is used to determine the intra- and inter-sublattice exchange constants and from the spin-flop transition we infer the anisotropy of the electrically active sublattice to be \SI{216}{\kilo\joule\per\cubic\meter} and predict the magnetic resonances frequencies.
Exchange and anisotropy are comparable and hard-axis applied magnetic fields result in a tilting of the magnetic moments from their collinear ground state.
Our analysis is applicable to collinear ferrimagnetic half-metal systems.

\end{abstract}

\maketitle

\onecolumngrid
\begin{center}

\textbf{PhySH:} Ferrimagnetism, Magnetotransport, Half-metals, Anomalous Hall effect, Magnetic anisotropy, Exchange interaction \linebreak

\end{center}

\twocolumngrid

Thin films with ultra-high magnetic anisotropy fields exhibit magnetic resonances in the range of hundreds of GHz \citep{Mizukami2011, Awari2016, Mizukami2016} which is promising for future telecommunications applications. 
Spin-transfer driven nano-oscillators (STNOs) working on the principle of angular momentum transfer from a spin-polarised current to a small magnetic element \cite{SLO96, BER96}, have achieved output powers of several \SI{}{\micro\watt} and frequency tuneabilities of $\sim$\SI{}{\giga\hertz\per\milli\ampere} \cite{Deac2008803, Ikeda2008}, useful for wireless data transmission  \cite{Choi2014}. 
Output frequencies of STNOs based on standard transition-metal based ferromagnets, such as CoFeB, or cubic Heulser alloys such as Co$_2$Fe$_{0.4}$Mn$_{0.6}$Si are in the low GHz range \cite{Ikeda2010, Rippard2010, Skowronski2012, Zeng2013, Yamamoto2015}. 

Certain Heusler alloys \cite{Graf20111, Ma2012} are a suitable choice for achieving much higher output frequencies, aimed at enabling communication networks beyond 5G \cite{Dhillon2017}. 
The Mn$_{3-x}$Ga family contains two Mn sublattices which are antiferromagnetically coupled in a ferrimagnetic structure \cite{Graf20111}. 
They have low net magnetization, $M_{\text{net}}$, and high effective magnetic anisotropy, $K_{\text{eff}}$, with anisotropy fields of $\upmu_0H_\text{K}=2K_{\text{eff}}/M_{\text{net}}$ exceeding \SI{18}{\tesla} \citep{Kurt2011solidi, Fowley2015}, which results in resonance frequencies two orders of magnitude higher \cite{Mizukami2011, Awari2016} than Co-Fe-B.
Furthermore, the magnetic properties of these ferrimagnetic alloys can be tuned easily with composition \cite{Glas2013134, Kurt2014, BET15}.
Mn$_{3-x}$Ga films have shown tuneable resonance frequencies between \SIrange{200}{360}{\giga\hertz} by variation of the alloy stoichiometry and magnetic anisotropy field \citep{Awari2016}. 

\begin{figure}[b!]
 \includegraphics[width=1\columnwidth, trim=0 0 0 0, clip]{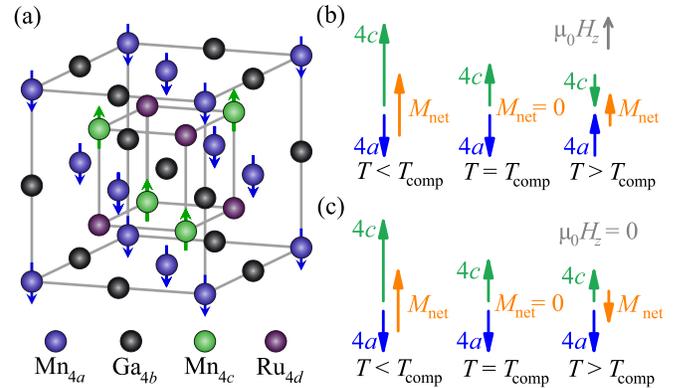}
 \caption{(a) Crystal structure of Mn$_2$Ru$_x$Ga, the magnetic moments of the Mn$_{4a}$ and Mn$_{4c}$ are aligned antiparallel. (b) and (c), a two-sublattice macrospin model used to explain the observed temperature and field dependences of electronic transport in the presence and absence of an applied field $\upmu_0H_{z}$, respectively. Two key points of the model are: below (above) $T_{\text{comp}}$ the moment of the Mn$_{4c}$ is parallel (antiparallel) to $M_\text{{net}}$; and, in the absence of an applied field the sublattice moments do not change their orientation upon crossing $T_{\text{comp}}$.
}
 \label{structure}
  \label{simplemodel}
\end{figure}

\begin{figure*}[t!]
 \includegraphics[width=\textwidth, trim=30 50 10 20, clip]{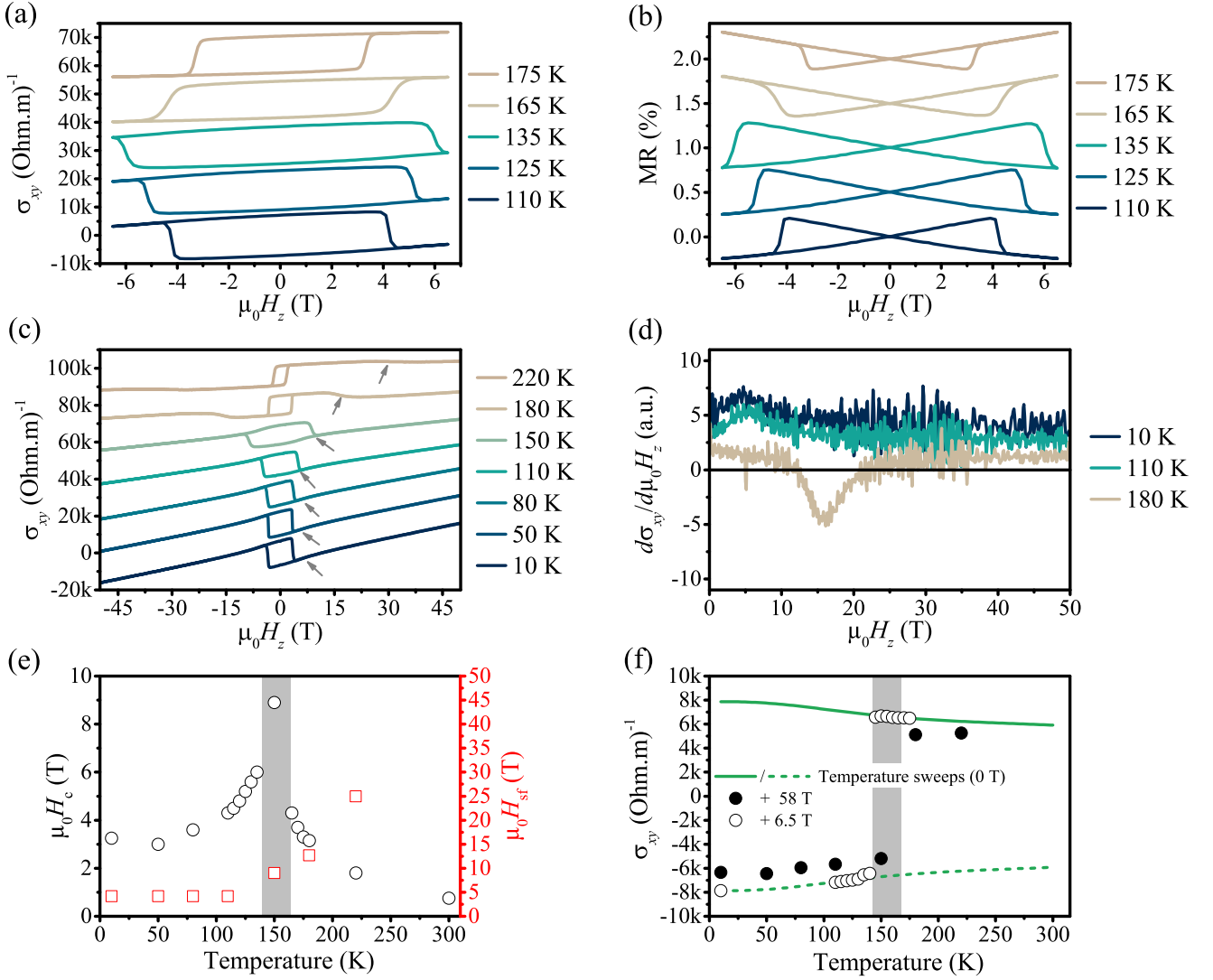}
 \caption{(a) AHC loops up to \SI{6.5}{\tesla} for Mn$_2$Ru$_{0.61}$Ga around the compensation temperature (\SI{155}{\kelvin}). Loops are offset vertically for clarity.
 (b) Magnetoresistance loops recorded at the same time as the data in (a). Loops are offset vertically for clarity.
 (c) AHC loops up to \SI{58}{\tesla}, where the spin-flop transition is indicated by the grey arrows. The linear slope is due to the ordinary Hall effect. Loops are offset vertically for clarity.
 (d) Derivative of the selected data in (c) clearly highlighting the spin-flop.
 (e) $\upmu_0 H_{\text{c}}$ (black circles) and $\upmu_0 H_{\text{sf}}$ (red squares) as a function of temperature. The divergence of the coercivity is expected at $T_{\text{comp}}$ since with $M_{\text{net}}=0$ and $K_{\text{eff}} \neq 0$. 
 (f) Temperature dependence of the remanent Hall conductivity when saturated at \SI{10}{\kelvin} in negative (solid line) and positive (dashed line) applied field. The black open (closed) circles record the remanent Hall resistivity after the application of \SI{6.5}{\tesla} (\SI{58}{\tesla}).}
 \label{low-field-ehe}
\end{figure*}

Here we focus on the fully-compensated half-metallic Heusler compound, Mn$_2$Ru$_x$Ga (MRG) \cite{Kurt2014,BET15,BOR16,THI15, karsten-unpub, Zic2016}.
Films of MRG were first shown experimentally \cite{Kurt2014} and subsequently confirmed by DFT calculations \cite{Zic2016} to exhibit a spin gap at $E_\text{F}$.
The material crystallises in the cubic space group, $F\bar{\text{4}}\text{3}m$.
Mn on the 4$a$ and 4$c$ sites are antiferromagnetically coupled, while those on the same sites are ferromagnetically coupled. 
The crystal structure is shown in figure \ref{structure} (a).
The Ga is on the 4$b$ sites and Ru occupies a fraction of the 4$d$ sites \cite{Kurt2014}.
We will discuss Mn on the 4$a$ and 4$c$ sites by referring to the Mn$_{4a}$ and Mn$_{4c}$ sublattices. 
By changing the Ru concentration, the magnetic properties of the Mn$_{4c}$ sublattice are altered, while those of the Mn$_{4a}$ sublattice remain relatively constant \cite{BET15}.
Thin films grown on MgO have an out-of-plane magnetic easy axis due to biaxial strain induced by the substrate during growth \cite{THI15}.
Unlike the uncompensated tetragonal $D$0$_{22}$ Mn$_{3-x}$Ga family of alloys, MRG has a compensation temperature, $T_{\text{comp}}$, where there is no net magnetization \cite{Kurt2014, BET15}.
Nonetheless, there is non-vanishing tunnel magnetoresistance \cite{BOR16}, spin Hall angle \cite{THI15} and magneto-optical Kerr effect \cite{karsten-unpub}, which all arise from the Mn$_{4c}$ sublattice. 
The occupied electronic states originating from the Mn$_{4a}$ sublattice lie below the spin gap \cite{Zic2016}.

The electrical transport on MRG reported to date \cite{THI15, BOR16} can be explained using the model shown in figure \ref{structure} (b) and (c) where the direction of spin polarisation is governed by the direction of the Mn$_{4c}$ sublattice and not Mn$_{4a}$ or $M_{\text{net}}$.
Here we make use of the dominant influence of a single sublattice on the electron transport to study the magnetism of a compensated half-metal at compensation, and evaluate the exchange and anisotropy energies.

We measure magnetotransport, especially the anomalous Hall effect in the temperature range \SIrange{10}{300}{\kelvin} in magnetic fields up to \SI{58}{\tesla}. 
The anomalous Hall conductivity (AHC) of a metallic ferromagnetic film, $\sigma_{xy}$, is proportional to the out-of-plane component of magnetization, $M_z$, which is defined as $M \cos \theta$ where $\theta$ is the angle between the $z$-axis and the magnetization, $M$ \cite{Nagaosa20101539}. 
In ferrimagnets, however, the AHC will depend on the band structure at the Fermi level, $E_{\text{F}}$, so when the material is half-metallic, one expects $\sigma_{xy} \propto M_{\text{sl}} \cos\theta_{M_{\text{sl}}}$, where $M_{\text{sl}}$ is the magnetization of the sublattice that dominates the transport. 

Mn$_2$Ru$_x$Ga layers of varying composition, $x$ = 0.55, 0.61 and 0.70, were deposited on MgO substrates in a fully automated Shamrock sputtering system.
The thickness of the films, $\approx$ \SI{27}{\nano\meter}, was determined by X-ray reflectivity. 
Hall crosses of width \SI{100}{\micro\meter} and length \SI{900}{\micro\meter} were patterned using direct-laser-write lithography, Ar$^+$ ion milling and lift-off. 
The Hall bars were contacted with Cr \SI{5}{\nano\meter} / Au \SI{125}{\nano\meter} pads.  

A Lakeshore Hall system was used to measure the longitudinal ($\rho_{xx}$) and transverse ($\rho_{xy}$) resistivities from \SIrange{10}{300}{\kelvin} in out-of-plane fields up to \SI{6.5}{\tesla}. 
The AHC, $\sigma_{xy} = \rho_{xy}/\rho^2_{xx}$ \citep{ZengPRL2006, LeeScience2004}, is obtained from the raw data.
In-plane, $\upmu_0 H_x$, and out-of-plane, $\upmu_0 H_z$, pulsed magnetic fields of up to \SI{58}{\tesla} were applied at the Dresden High Magnetic Field Laboratory at selected temperatures between \SI{10}{\kelvin} and \SI{220}{\kelvin}.  
We focus on Mn$_2$Ru$_{0.61}$Ga with $T_{\text{comp}} \approx$ \SI{155}{\kelvin}.
All three compositions were found to have compensation temperatures between \SI{100}{\kelvin} and \SI{300}{\kelvin}, and exhibit similar properties. 

AHC loops versus $\upmu_0 H_z$ around $T_{\text{comp}}$ are shown in figure \ref{low-field-ehe} (a).
At all temperatures, MRG exhibits strong perpendicular magnetic anisotropy. 
The reversal of the sign of $\sigma_{xy}$ between \SI{135}{\kelvin} and \SI{165}{\kelvin} indicates a reversal of the spin polarisation at $E_{\text{F}}$ with respect to the applied field direction, as expected on crossing $T_{\text{comp}}$. 
The coercivity, $\upmu_0 H_{\text{c}}$, varies from \SIrange{3}{6}{\tesla} between \SI{110}{\kelvin} and \SI{175}{\kelvin}.
The longitudinal magnetoresistance, $\rho_{xx}(H) / \rho_{xx}(0)$ shown in figure \ref{low-field-ehe} (b) is small (\textless 1\%), as expected for a half-metal \cite{Yang_AMR_PRB_2012}.
Pulsed field measurements in figure \ref{low-field-ehe} (c) show that, close to $T_{\text{comp}}$, $\upmu_0 H_{\text{c}}$ exceeds \SI{9}{\tesla} and that MRG exhibits a spin-flop transition at higher fields, indicated in the figure by the grey arrows.
The derivative of selected curves of $\sigma_{xy}$ versus applied field, figure \ref{low-field-ehe} (d), show clearly the spin-flop field especially at lower temperatures.
We note that the longitudinal magnetoresistance up to \SI{58}{\tesla} also does not exceed 1\% (not shown).
The divergence in coercivity (black circles in figure \ref{low-field-ehe} (e)) is expected at $T_{\text{comp}}$ because the anisotropy field in uniaxial magnets is $\upmu_0H_\text{K}=2K_{\text{eff}}/M_{\text{net}}$, where $K_{\text{eff}}$ is the effective anisotropy energy and $M_{\text{net}}$ is the net magnetization. 
The anisotropy field is an upper limit on coercivity. 
The temperature dependence of the spin-flop field, $\upmu_0 H_{\text{sf}}$, is also plotted in figure \ref{low-field-ehe} (e) (red squares).

The solid (dashed) line in figure \ref{low-field-ehe} (f), traces the temperature dependence of $\sigma_{xy}$ when the sample is initially saturated in a field of \SI{-6.5}{\tesla} (\SI{+6.5}{\tesla}) at \SI{10}{\kelvin} and allowed to warm up in zero-applied magnetic field. 
The spontaneous Hall conductivity, $\sigma_{xy}$, decreases from \SIrange{7859}{5290}{\per\ohm\per\meter} and does not change sign for either of the zero-field temperature scans. 
The remanent value of $\sigma_{xy}$ after the application of \SI{6.5}{\tesla} (\SI{58}{\tesla}) is plotted with open (solid) symbols. 
The combined data establish that, in MRG films, neither the AHE nor the AHC are proportional to $M_{\text{net}}$. 
They depend on the magnetization of the sublattice that gives rise to $\sigma_{xy}$.
While similar behaviour is well documented for the anomalous Hall effect in rare-earth – transition-metal (RE-TM) ferrimagnets, where both RE and TM elements contribute to the transport \cite{Ogawa197687, Mimura1976, McGuire1977}, in MRG both magnetic sublattices are composed of Mn which has been confirmed to have the same electronic configuration, 3$d^{5}$ \cite{BET15}.
If both sublattices contributed equally to the effect, the sum should fall to zero at $T_{\text{comp}}$. 

We refer to the model presented in figure \ref{structure} (b) and (c) to explain the behaviour shown in figure \ref{low-field-ehe} (f). 
Figure \ref{simplemodel} (b) shows the Mn$_{4a}$ and Mn$_{4c}$ sublattice moments and the net magnetic moment in the case of an applied field, $\upmu_0 H_z$, along the easy-axis of MRG.
Below $T_{\text{comp}}$, the Mn$_{4c}$ moment (green arrow) outweighs that of Mn$_{4a}$ (blue arrow), and $M_{\text{net}}$ (orange arrow) is parallel to the Mn$_{4c}$ sublattice. 
At $T_{\text{comp}}$, $M_{\text{net}}$ is zero but the directions of the sublattice moments have not changed with respect to $\upmu_0 H_z$. 
Above $T_{\text{comp}}$, $\upmu_0 H_z$ causes a reversal of $M_{\text{net}}$ (provided it exceeds $\upmu_0 H_{\text{c}}$).
Here, the Mn$_{4a}$ sublattice has a larger moment than Mn$_{4c}$ and $M_{\text{net}}$ will be in the same direction as the Mn$_{4a}$ moment.
Due to the antiferromagnetic alignment of both sublattices the moment on Mn$_{4c}$ is parallel (antiparallel) to $\upmu_0 H_z$ below (above) $T_{\text{comp}}$. 

In the absence of an applied field (figure \ref{simplemodel} (c)), the direction of $M_{\text{net}}$ will reverse crossing $T_{\text{comp}}$ due to the different temperature dependences of the sublattice moments. 
However, the sublattice moments only change in magnitude, and not direction.
The uniaxial anisotropy provided by the slight substrate-induced distortion of the cubic cell \cite{Kurt2014} provides directional stability along the $z$-axis.
Therefore, crossing $T_{\text{comp}}$ in the absence of applied field, we expect no change in sign of $\sigma_{xy}$, nor should it vanish.
The Mn$_{4c}$ sublattice dominates the electron transport and determines spin direction of the available states at $E_{\text{F}}$, while the Mn$_{4a}$ states form the spin-gap.

\begin{figure}[t!]
 \includegraphics[width=1\columnwidth, trim=33 50 40 20, clip]{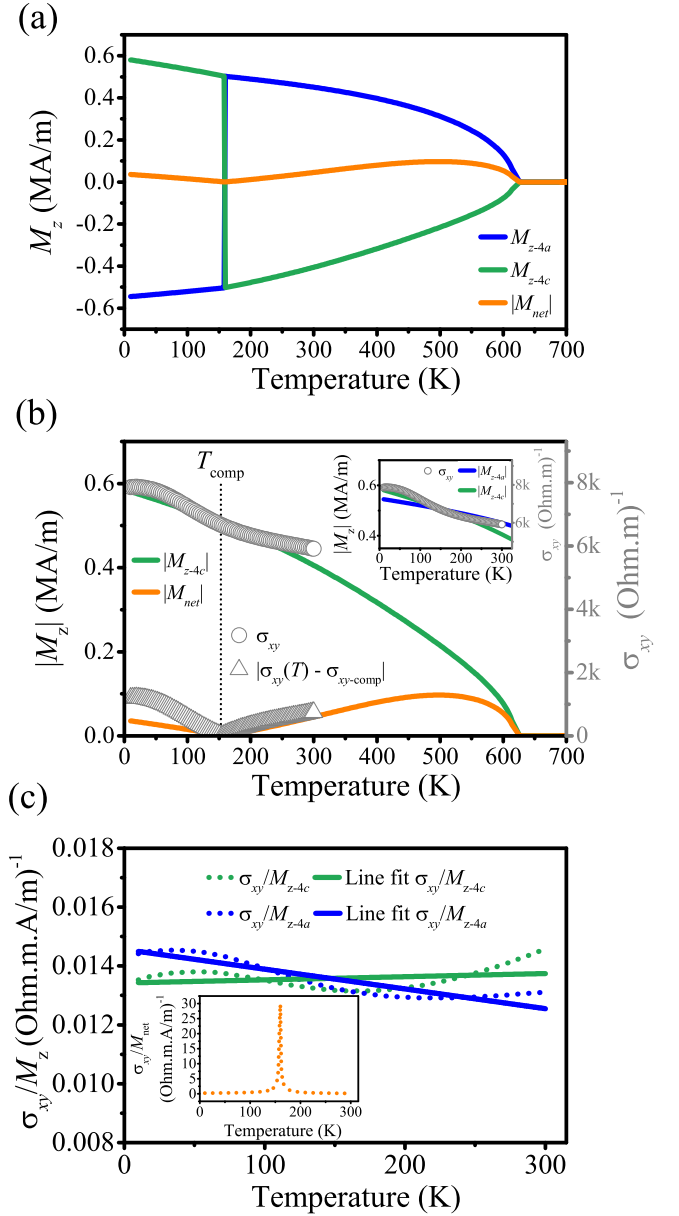}
 \caption{
 (a) Temperature dependence of $M_{z-4a}$ and $M_{z-4c}$ and $M_{\text{net}}$. The data is obtained from numerical integration with an applied field to set the direction of the net magnetization and then reduced to zero, therefore the magnetization reverses at $T_{\text{comp}}$. $T_{\text{comp}}$ is \SI{155}{\kelvin} and the Curie temperature is \SI{625}{\kelvin}. 
 (b) $\sigma_{xy}$, $\lvert \sigma_{xy} (T) -\sigma_{xy-\text{comp}}\rvert$, $\lvert M_{z-4c}\rvert$, and $\lvert M_{\text{net}} \rvert$ as a function of temperature. Inset: $\sigma_{xy}$ plotted with $M_{z-4a}$ and $M_{z-4c}$ as a function of temperature, to show that $\sigma_{xy}$ does indeed follow $M_{z-4c}$ and not $M_{z-4a}$.
 (c) Ratio of $\sigma_{xy} / M_{z-4c}$ and $\sigma_{xy} / M_{z-4a}$ (dotted lines) over the experimentally measured temperature range complete with linear fits (solid lines). The ratio is almost constant with no significant linear background slope showing that $\sigma_{xy} \propto M_{z-4c}$. The inset shows the clear divergence of $\sigma_{xy} / M_{net}$ at $T_{\text{comp}}$.}
 \label{model}
\end{figure}

The results of a molecular field model \cite{smart1966effective} based on two sublattices are presented in figure \ref{model}.
The molecular field, $\textbf{\emph{H}}^{i}$, experienced by each sublattice is given by:
\begin{equation}
\textbf{\emph{H}}^{i}_{4a} = n_{4a-4a} \textbf{\emph{M}}_{4a} + n_{4a-4c} \textbf{\emph{M}}_{4c} + \textbf{\emph{H}}
\label{eqn1}
\end{equation}
\begin{equation}
 \textbf{\emph{H}}^{i}_{4c} = n_{4a-4c} \textbf{\emph{M}}_{4a} + n_{4c-4c} \textbf{\emph{M}}_{4c} + \textbf{\emph{H}}
\label{eqn2}
\end{equation}

where $n_{4a-4a}$ and $n_{4c-4c}$ are the intra-layer exchange constants and $n_{4a-4c}$ is the inter-layer exchange constant.
$\textbf{\emph{M}}_{4a}$ and $\textbf{\emph{M}}_{4c}$ are the magnetizations of the 4$a$ and 4$c$ sublattices.
$\textbf{\emph{H}}$ is the externally applied magnetic field.
The moments within the Mn$_{4a}$ and Mn$_{4c}$ sublattices are ferromagnetically coupled and hence $n_{4a-4a}$ and $n_{4c-4c}$ are both positive.
The two sublattices couple antiferromagnetically and therefore $n_{4a-4c}$ is negative.
The equations are solved numerically for both temperature and applied field dependences to obtain the projection of both sublattice magnetizations along the $z$-axis, $M_{z-\alpha} = M_{\alpha} \cos \theta_{\alpha} $, where $\alpha = 4a, 4c$.
%, which in the end is proportional to $\sigma_{xy}$.
In the absence of an applied field, $\theta=0$, therefore $M_{z-\alpha}$ reduces simply to $M_{\alpha}$.

\begin{table}[!b]
\caption{Initial parameters input to the molecular field model according to equations \ref{eqn1} and \ref{eqn2}. $M_{4a}$, $M_{4c}$ and $K_{4a}$, $K_{4c}$ are the magnetizations and uniaxial anisotropies on the 4$a$, 4$c$ sublattices. $n_{4a-4a}$ and $n_{4c-4c}$ are the intralayer exchange constants. $n_{4a-4c}$ is the interlayer exchange constant. Derived parameters are outputs of the molecular field model. \label{tab:table}} 
\begin{ruledtabular}
\begin{tabular}{cccc}
\multicolumn{4}{  c  }{Initial parameters} \\ \midrule
$M_{4a}$ (\SI{0}{\kelvin}) & \SI{547}{\kilo\ampere\per\meter} & $n_{4a-4a}$ & 1150 \\
$M_{4c}$ (\SI{0}{\kelvin}) & \SI{585}{\kilo\ampere\per\meter} & $n_{4c-4c}$ & 400  \\ 
$K_{4a}$ & \SI{0}{\kilo\joule\per\cubic\meter} & $n_{4a-4c}$ & -485 \\ 
$K_{4c}$ & \SI{216}{\kilo\joule\per\cubic\meter} & & \\\midrule
\multicolumn{4}{ c }{Derived parameters} \\ \midrule
$M_{\text{net}}$ (\SI{10}{\kelvin}) & \SI{38}{\kilo\ampere\per\meter} & $T_{\text{C}}$ & \SI{625}{\kelvin}  \\ 
$M_{\text{net}}$ (max.) & \SI{97}{\kilo\ampere\per\meter} &  $T_{\text{comp}}$ & \SI{155}{\kelvin} \\
\end{tabular}  
\end{ruledtabular}
\end{table}

The model parameters are given in table \ref{tab:table}.
Based on previous XMCD measurements \cite{BET15} as well as DFT calculations \cite{Zic2016} we take values of \SI{547}{\kilo\ampere\per\meter} and \SI{585}{\kilo\ampere\per\meter} for the magnetizations on the 4$a$ and 4$c$ sublattice, respectively.
The values of $n_{4a-4a}$, $n_{4c-4c}$ and $n_{4a-4c}$ are fit to reproduce $T_{\text{comp}}$ and the Curie temperature, $T_{\text{C}}$.
The temperature dependences of $M_{z-4a}$ (blue line), $M_{z-4c}$ (green line) and $M_{\text{net}}$ (orange line) with $n_{4a-4a} = 1150 $, $n_{4c-4c} = 400 $ and $n_{4a-4c} = -485 $ are shown in figure \ref{model} (a). 
In order to numerically obtain the temperature dependence in zero applied field a strong field of \SI{60}{\tesla} is used to set the direction of $M_\text{net}$ and then reduced to zero, so the sublattice moments reverse at $T_{\text{comp}}$ = \SI{155}{\kelvin} as in the experiment.
$T_{\text{C}}$ is \SI{625}{\kelvin}.
$M_{\text{net}}$ varies from \SI{38}{\kilo\ampere\per\meter} at \SI{10}{\kelvin} to a maximum of \SI{97}{\kilo\ampere\per\meter} at \SI{512}{\kelvin}, close to $T_{\text{C}}$.

Figure \ref{model} (b) shows the measured AHC (circles), along with $\lvert M_{z-4c}\rvert$ (green line) from the molecular field model.
It can be seen clearly that $\sigma_{xy}$ follows the temperature dependence of $M_{z-4c}$ below $T_{\text{comp}}$ and not $M_{\text{net}}$. 
As a further step, we plot $\lvert M_{\text{net}}\rvert$ from the molecular field model (orange line) with $\lvert \sigma_{xy}(T) - \sigma_{xy-comp}\rvert$ (triangles). 
As $\sigma_{xy}$ is proportional only to $M_{4c}$ and at compensation $M_{4c} = M_{4a}$, subtracting the value of $\sigma_{xy}$ at $T_{\text{comp}}$ ($|\sigma_{xy}(T) - \sigma_{xy-comp}|$) gives an approximate indication of how $M_{\text{net}}$ behaves with temperature. 
Even though this ignores the weak $M_{4a}$ temperature dependence, the trend of $M_{\text{net}}$ follows $|\sigma_{xy}(T) - \sigma_{xy-comp}|$, showing that $\sigma_{xy}$ is a reflection of $M_{4c}$ and not $M_{\text{net}}$. 
The inset in figure \ref{model} (b) shows both $M_{z-4a}$ and $M_{z-4c}$ with the experimentally obtained $\sigma_{xy}$, and shows that $\sigma_{xy}$ more closely follows $M_{z-4c}$.
Figure \ref{model} (c) shows the ratio of $\sigma_{xy}$ to $M_{z-4c}$ (green dotted line), $M_{z-4a}$ (blue dotted line) and $M_{\text{net}}$ (inset).
Linear fitting of $\sigma_{xy}/M_{z-4c}$ (solid green line), and $\sigma_{xy}/M_{z-4a}$ (solid blue line) show that $\sigma_{xy}/M_{z-4c}$ remains constant over the measured temperature range, and is equal to \SI{0.0136}{\per\ohm\per\meter\per\ampere\meter}, similar to what has been reported for other itinerant ferromagnetic systems \cite{Manyala2004255,HusmannSinghAHC}.
The linear slope for $\sigma_{xy}/M_{z-4a}$ and the divergence of $\sigma_{xy}/M_{\text{net}}$ shows that $\sigma_{xy}$ reflects neither of these two quantities.

A recent study has shown via \emph{ab initio} calculations that this must be the case for a fully-compensated half-metallic ferrimagnetic system \cite{Stinshoff2017-ArXiv} although previous reports on bulk films found $\rho_{xy}$, and hence, $\sigma_{xy}$ falling to zero at $T_{\text{comp}}$ \cite{Stinshoff2017}. 

For the evaluation of the magnetic anisotropy we use the initial low-field change of $\sigma_{xy}$ versus $\upmu_0 H_x$ and extrapolate to zero and obtain $K_{4c}$ (not shown). 
The values obtained vary from \SIrange{100}{250}{\kilo\joule\per\cubic\meter} over the entire data range.
We also calculate the anisotropy directly from the spin-flop transition $H_{\text{sf}} = \sqrt{2 H_{\text{K}} H^{ex}_{4c}}$, where $H_{\text{K}}$ is the \emph{sublattice} anisotropy field and $H^{ex}_{4c}$ is the exchange field, the first term in Eqn. \ref{eqn2}.
The anisotropy field, $H_{\text{K}}$, is related to the sublattice anisotropy energy $K_{4c}$.

\begin{figure}
\includegraphics[width=1\columnwidth, trim=0 5 0 0, clip]{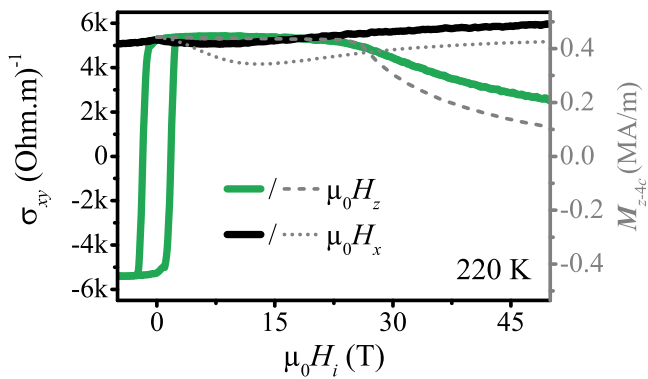}
\caption{Comparison of experimental data (solid lines) and molecular field model (dashed lines) at \SI{220}{\kelvin} for fields applied along $\upmu_0 H_z$ and $\upmu_0 H_x$. $\upmu_0 H_{\text{sf}}$ is observed in both cases at \SI{26}{\tesla}.}
\label{finalfig}
\end{figure}

A comparison between the experiment and the model at \SI{220}{\kelvin} for both $\upmu_0 H_z$ and $\upmu_0 H_x$ is shown in figure \ref{finalfig}.
The solid lines plot the experimentally obtained $\sigma_{xy}$, while the dashed lines plot $M_{z-4c}$ from the model. 
The spin-flop field is observed in both cases at $\upmu_0 H_z$ = \SI{26}{\tesla}.
For the case of $\upmu_0 H_x$, it can first be seen that the 4$c$ moment does not saturate along the field as one would expect \cite{Fowley2015, Guo2006}. 
It initially decreases but then returns to a saturated value in both the experimental data and the model. 
This behaviour is due to the fact that in MRG the exchange and anisotropy energies are comparable and weak. 
If the exchange coupling is strong then the net magnetic moment could be saturated along $\upmu_0 H_x$ as both sublattices can remain antiparallel up to the anisotropy field $\upmu_0 H_\text{K} = 2 (K_{\text{4}a}+K_{\text{4}c}) / (M_{\text{4}a} +M_{\text{4}c}) = 2 K_{\text{eff}} / M_{\text{net}}$.
If the exchange coupling is weak then both sublattice moments will tilt from their antiparallel alignment, breaking exchange, before the net magnetic moment can be saturated along $\upmu_0 H_x$ at the appropriate sublattice anisotropy field $\upmu_0 H_{\text{K}} = 2 K_{\text{sl}} / M_{\text{sl}}$, $\text{sl} = \text{4}a, \text{4}c$.

The model and experiment disagree slightly on the temperature dependence of $H_{\text{sf}}$ below $T_{\text{comp}}$. 
Better agreement can be obtained by using much higher anisotropy energies of opposite sign: $K_{4a} = \SI{-1.5}{\mega\joule\per\cubic\meter}$ and $K_{4c} = \SI{1.7}{\mega\joule\per\cubic\meter}$. 
This has the effect of increasing (decreasing) $H_{\text{sf}}$ above (below) $T_{\text{comp}}$.
While this improves the match between $\sigma_{xy}$ versus $\upmu_0 H_z$ below $T_{\text{comp}}$, it worsens the match of $\sigma_{xy}$ versus $\upmu_0 H_x$, at all temperatures.
This and the slight discrepancies between the model and experiment when a low value of $K_{4c}$ is used (Fig. \ref{finalfig}) indicate that additional anisotropies, likely cubic, in MRG, as well as anti-symmetric exchange (Dzyaloshinskii-Moriya interaction) should be taken into account.

We have shown that the uniaxial molecular field model reproduces the main characteristics of the experimental data and we confirm the relationship $\sigma_{xy} \propto M_{4c} \cos\theta_{M_{4c}}$.
Knowing $H_{\text{K}}$ and $H^{ex}_{4c}$ we can predict the frequencies of the anisotropy, $f_{anis} = \gamma \upmu_0H_{\text{K}}$, and the exchange, $f_{exch} = \gamma\upmu_0\sqrt{2 H_{\text{K}}  H^{ex}_{4c}} = \gamma\upmu_0 H_{\text{sf}}$, magnetic resonance modes, where $\gamma = \SI{28.02}{\giga\hertz\per\tesla}$ \cite{moorish}.
At \SI{220}{\kelvin}, $\upmu_0 H_{\text{sf}}$ = \SI{26}{\tesla} and $\upmu_0 H^{ex}_{4c} = n_{4a-4c}M_{4a}$ = \SI{294}{\tesla}, therefore $\upmu_0 H_{\text{K}}$ = \SI{1.15}{\tesla} and the resonances are $f_{anis} = \SI{32}{\giga\hertz}$ and $f_{exch} = \SI{729}{\giga\hertz}$.

In conclusion, $\sigma_{xy}$ for fully-compensated half-metallic ferrimagnetic alloys follows the relevant sublattice magnetization, $M_{\text{sl}} \cos\theta_{M_{\text{sl}}}$, and not $M_{\text{net}} \cos\theta_{M_{\text{net}}}$. 
High-field magnetotransport and molecular field modelling allows the determination of the anisotropy and exchange constants provided the half-metallic material is collinear.  
Mn$_2$Ru$_x$Ga behaves magnetically as an antiferromagnet and electrically as a highly spin polarised ferromagnet; it is capable of operation in the \SI{}{\tera\hertz} regime and its transport behaviour is governed by the Mn$_{4c}$ sublattice.
The immediate, technologically relevant, implication of these results is that spin-transfer torque effects in compensated ferrimagnetic half-metals will be governed by single sublattice.

\section{Acknowledgements}

This project has received funding from the European Union’s Horizon 2020 research and innovation programme under grant agreement No 737038 (TRANSPIRE). This work is supported by the Helmholtz Young Investigator Initiative Grant No. VH-N6-1048. We acknowledge the support of the HLD at HZDR, member of the European Magnetic Field Laboratory (EMFL).

\end{document}